\documentclass[twocolumn,showpacs,preprintnumbers,amsmath,amssymb]{revtex4}
\usepackage{graphicx}
\usepackage{dcolumn}
\usepackage{bm}

\renewcommand{\deg}{{}^{\circ}}

\begin{document}
\title{Erosion waves: transverse instabilities and fingering.}
\author{Florent Malloggi}
\author{Jos\'{e} Lanuza}
\author{Bruno Andreotti}
\author{Eric Cl\'{e}ment}
\affiliation{Laboratoire de Physique et M\'ecanique des Milieux H\'et\'erog\`enes, 10 rue Vauquelin 75005 Paris France, UMR CNRS 7636}
\date{\today}

\begin{abstract}
Two laboratory scale experiments of dry and under-water avalanches of non-cohesive granular materials are investigated.  We trigger solitary waves and study the conditions under which the front is transversally stable. We show the existence of a linear instability followed by a coarsening dynamics and finally the onset of a fingering pattern. Due to the different operating conditions, both experiments strongly differ by the spatial and time scales involved. Nevertheless, the quantitative agreement between the stability diagram, the wavelengths selected and the avalanche morphology reveals a common scenario for an erosion/deposition process.
\end{abstract}
\pacs{47, 47.10.+g,  68.08.-p, 68.08.Bc}

\maketitle
Avalanching processes leading to catastrophic transport of various natural materials do not only occur in the air as we know of snow avalanches, mud flows and their catastrophic human and economical toll. Such events frequently happen below the see level as they take many forms from turbidity currents to thick sediment waves sliding down the continental shelf. This is a fundamental feature shaping the submarine morphology. From the modeling of risks point of view, important questions still remain such as to evaluate to which extend an initial triggering event (an earth quake, an eruption..) would be responsible for a subsequent process that might propagate or amplify over large distances as an unstable matter wave. Unfortunately, the dynamics of such catastrophic events remains an issue so far lacking of conceptual clarity \cite{I97,HLL96} since (i) the rheology of the flows involved in an avalanche is complex and still unraveled, (ii) the physics of erosion/deposition mechanisms is essentially limited to empirical descriptions based on dimensional analysis and semi-empirical formulations.
\begin{figure}[t!]
\includegraphics{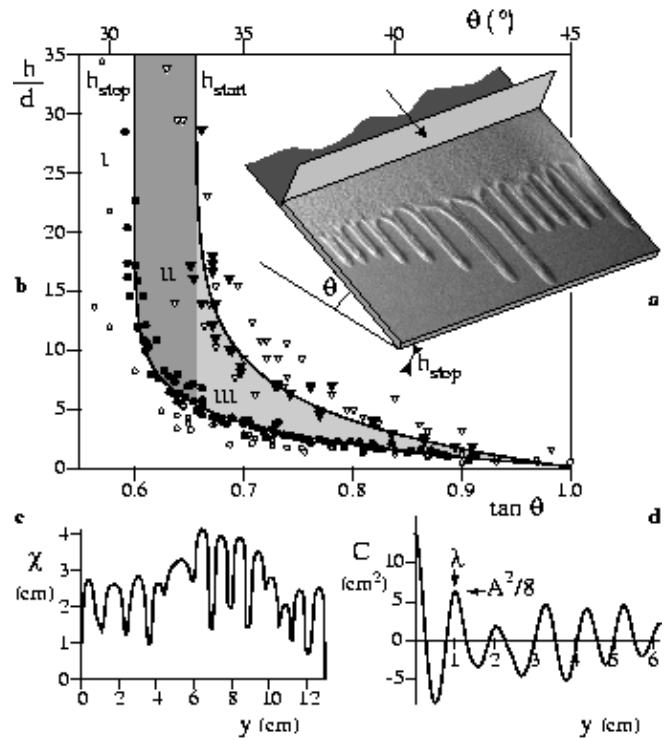}
\caption{{\bf a} Experimental set-up. {\bf b} Stability diagram: $h_{stop}$ is the thickness of the sediment left after an avalanche for a given angle $\theta$, in air ({\Large $\bullet$}) and in water ({\Large $\circ$}); $h_{start}(\theta)$ is the maximum stable height of sediment, in air ($\blacktriangledown$) and in water ($\triangledown$). $h_{start}(\theta)$ and $h_{stop}(\theta)$ are fitted by the form $h=b \ln((\tan \theta - \mu)/\delta \mu)$ (solid lines). Avalanches triggered in region II  are stable while they exhibit a transverse instability in region III.  In particular, solitary erosion waves are evidenced when starting from the stable height $h_{stop}$. {\bf c} Front profile $\chi(y)$ obtained after image processing by a correlation technique.  {\bf d} The corresponding correlation function $C(y)$ allows to define the average wavelength $\lambda$ and amplitude $A$.}
\label{setup}
\end{figure}

Recently, extensive laboratory-scale experiments on dry granular materials have aimed to unify  the rheology of dense particulate flows in different geometries \cite{GDR04}. These flows can be organized into two sub-classes, quasi-static creep flows characterised by an exponential-like velocity profile, and shear flows, whose rheology is local \cite{CERC05,EH02,P04}. In the case of fully developed granular flows, measurements are now sufficiently precise and reproducible to evidence dependencies of the constitutive relation on microscopic granular features like rough sandy material vs round spherical particles \cite{FP03,BHE05}. But still, a full understanding of the flow constitutive relations would require a pertinent description of the passage between the blocked and the mobile phases (the jamming transition). For instance, the onset and the dynamics of triangular shape avalanches was shown to be  controlled essentially by the presence of a metastable substrate and the interplay between erosion and deposition processes \cite{DD99,D01}. From a theoretical point of view, descriptions of the granular bed mobilisation have been proposed, either empirically (`\`a la Saint-Venant') \cite{BCPE94,DAD99} or phenomenologically by phase field methods \cite{VTA03}. 

Avalanche fronts flowing on solid rough substrates are transversally stable, the transverse coupling due to gravity being essentially a stabilizing mechanism \cite{P99,PV99}. But, in the particular case of a strong size bidispersity, an avalanche front may exhibit a fingering pattern explained by a pinning mechanism \cite{PDS97,GT04}: the grains of larger size gathering in the finger thrusts are hindering locally the avalanche front progression. In this paper, we present an experimental study of avalanche fronts developing over an erodible granular substrate, in the air and under water.  We demonstrate the existence of a linear transverse instability of solitary erosion waves, although the rough grains we use exhibit a narrow polydispersity.
\begin{figure}[t!]
\includegraphics{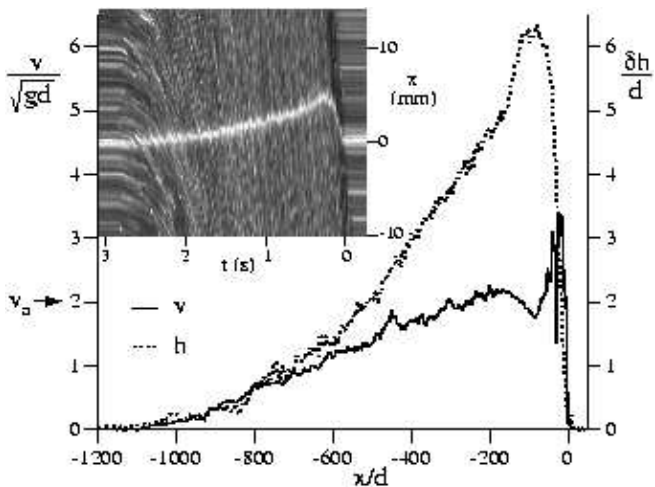}
\caption{Solitary erosion wave profile $\delta h=h-h_{stop}$ rescaled by $d$ (dotted line) and surface velocity profile $v$ rescaled by $\sqrt{gd}$ (solid line) (dry, $\theta =32\deg$, region II). Inset: spatio-temporal diagram done with a fast camera ($125$~Hz), showing the particle motion as well as the profile height (deflection of the laser sheet). It can be observed that the surface grain velocity tends at the front towards the solitary wave velocity $v_a$}
\label{onde}
\end{figure}

The avalanching set-ups consist of a thin layer of grains deposited on a substrate that can be tilted at a value $\theta $. The dry granular set-up is similar to the one of Daerr et al. \cite{DD99,D01}. The avalanche track is $70$~cm wide and $120$~cm long. The granular medium is Fontainebleau sand of a medium size $d=300~\mu$m and the track bottom is made of black velvet. For under water avalanches, the set up size is quite smaller. The avalanche track is the bottom of a plexiglass tank that can be tilted up to an angle $\theta$ from an horizontal position. The avalanche track width is $15$~cm and so is the track length. The granular sediment is an aluminum oxide powder of size either $d=30~\mu$m or $40~\mu$m. To avoid interparticle cohesion, it is sufficient to maintain the $pH$ value close to $4$ by adequate addition of hydrochloric acid \cite{D03}. The substrate is initially set at an horizontal position and a fixed mass of powder is poured and suspended by vigorous stirring. A uniform sediment layer of height $h$ then forms within $10$~min. The bottom is an abraded but transparent plexiglass plate which offers the possibility to monitor the avalanche dynamics by transparency when illuminated from below. The profile of the avalanche front $h(x,t)$ is obtained with a laser slicing technique and is resolved within $30~\mu$m ($0.1~d$). The front dynamics is quantitatively monitored by image processing of the avalanche front pictures. The front line equation  $\chi(y,t)$ is then extracted (fig.~\ref{setup}c) and the front line auto correlation function $C(y,t)=<x(y+y',t) x(y',t)>_{y'}$ is computed. Then, the correlation function first maximum is identified from which we define the average wavelength $\lambda$ and the amplitude $A=2\sqrt{2C(\lambda)}$ (fig.~\ref{setup}d). In addition, for dry avalanches, we measure the surface velocity field using a Particle Image Velocimetry technique.

It has been shown that the stability of dry granular layers of depth $h$ lying on a substrate inclined at an angle $\theta$ can be simply apprehended by a diagram with two branches \cite{P99} (fig.~\ref{setup}b) $h_{start}(\theta)$ and $h_{stop}(\theta )$ with the following interpretation:  a uniform deposit of height $h$ will globally loose stability if tilted above the angle $\theta$ defined by $h=h_{start}(\theta)$ and the avalanching process will leave at rest a deposit of height $h_{stop}(\theta)$. The $h_{start}$ and $h_{stop}$ curves diverge at  an asymptotic angle limit, respectively equal to the avalanche angle of the granular pile $\theta_{a}$ and to the repose angle $\theta_{r}$. Between the two, a domain of metastability for the granular deposit is present. Interestingly, the stability curves obtained for dry and underwater layers bear the same features and fall on the same curve when the deposited height is rescaled by the grain size (fig.~\ref{setup}b).
\begin{figure}[t!]
\includegraphics{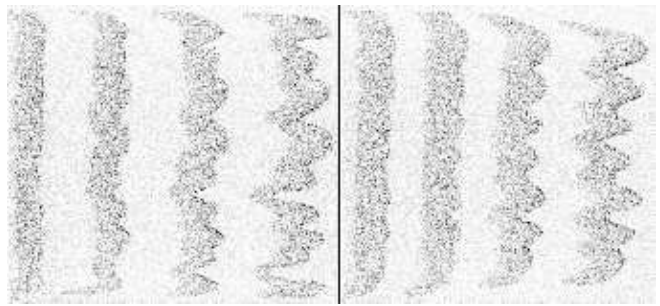}
\caption{Flowing part of solitary waves visualised by image difference (air,  $d=300~\mu$m, $\theta=35~\deg$, time interval $1.1$~s), starting from a flat bed (left) or from an initial bed presenting a forced wavelength $\lambda=6.5$~cm.}
\label{insta}
\end{figure}

To initiate avalanche fronts both in air and under water, we designed a `bulldozer' technique where a plate perpendicular to the avalanche track scrapes the sediment at a constant velocity (fig.~\ref{setup}a). Although our results on avalanche stability are valid in the whole metastable region (fig.~\ref{setup}b),
we will limit ourselves here to experiments  started from a stable sediment layer of height $h_{stop}(\theta )$. Once an autonomous avalanche front separates from the plate, the bulldozer driving stops.  We obtain both for dry and under water avalanches the following results. 

{\it (i)} For $\theta<\theta_{r}$ the sediment is quite stable as an avalanche front cannot propagate autonomously down the slope (region $I$ on graph 1b): the perturbation is bound to fade away when the driving stops. 

{\it (ii)} For $\theta_{r}<\theta <\theta _{a}$, we always obtain transversely stable autonomous avalanche fronts. We observed that the avalanche quickly converges toward a form which then remains constant. Furthermore, this solitary wave is found to be quite insensitive to the avalanche preparation details within a range of scraping velocities or initial masses set into motion. For this systematic study, we have kept a constant scraping velocity at about one-third of the typical avalanche velocity $v_a$. For each value of the -- unique -- control parameter $\theta$, there is thus a single possible  solitary erosion wave. In water, $v_a$ is of the order of the Stokes velocity $\dfrac{\Delta \rho} {\rho _{w}}\dfrac{gd^{2}}{18~\nu} \simeq 2$~mm/s where $\dfrac{\Delta \rho }{\rho _{w}}=3$ is the density contrast between grains and water, $\nu$ the water kinematic viscosity and $g$ the gravity acceleration. In the air, the propagation velocity is of the order of $\sqrt{gd}\simeq 5$~cm/s. In figure~\ref{onde} we show local sediment height $h$ and local surface velocity $v$ profiles for such an avalanche. It is important to note that the relation between $h$ and $v$ does not correspond to the curve found for homogeneous steady flows \cite{GDR04}. We rather obtain a relation of the form $v \propto \sqrt{g (h-h_{stop})}$. This result suggests that the grains are not flowing down to the bottom and consequently, erosion/deposition processes are determinant to select the avalanche dynamics and shape. 
\begin{figure}[t!]
\includegraphics{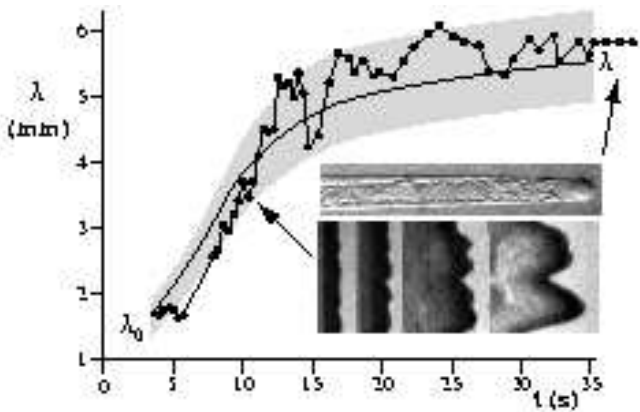}
\caption{Time evolution of the wavelength $\lambda$ (water, $d=40~\mu$m, $\theta=37.1\deg$) in a single typical realisation ({\Large $\bullet$})  and averaged  over realisations (solid line) -- the shadow zone indicates the standard deviation. After a small plateau at the initial wavelength $\lambda_0$, $\lambda$ increases due to merging processes (lower photograph) until the value $\lambda_\infty$ which corresponds to the formation of non-interacting fingers (upper photograph).}
\label{coarsening}
\end{figure}

\begin{figure}[t!]
\includegraphics{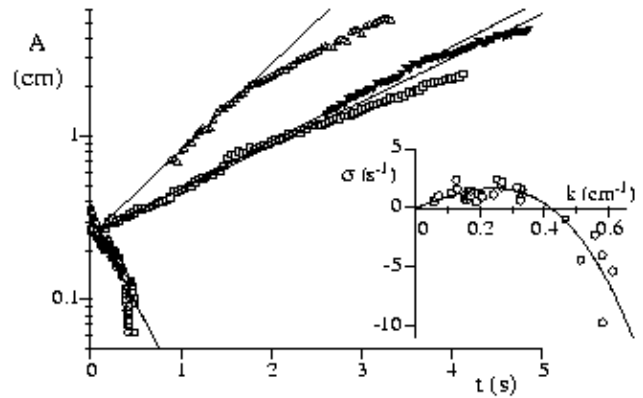}
\caption{Time evolution of the amplitude (air, $d=300~\mu$m, $\theta=35\deg$) when the initial condition is forced  at a given wavelength $\lambda=12~$mm ({\Large $\circ$}), $\lambda=30~$mm ($\square$),  $\lambda=90~$mm ($\triangle$) and $\lambda=178$~mm ($\blacktriangledown$). Inset: linear growth rate $\sigma$ as a function of the wave number $k$. The solid line is the best fit by $\sigma=\sigma_m |k|\lambda_0 (1-(k \lambda_0)^2/3)$, with a maximum growth rate $\sigma_m\simeq2.5~s^{-1}$ for $\lambda_0\simeq4$~cm.}
\label{instalin}
\end{figure}

{\it (iii)} For $\theta>\theta_{a}$ the neutral wave fronts are transversally unstable. After the initial instability, we have identified a sequence of fusion processes increasing the spatial modulation lengths (coarsening scenario). Finally, the transverse destabilization ends up as a fingering pattern. In this final stage, the flowing zones are disconnected one from the others so that the wavelength does not evolve anymore. On figure~\ref{coarsening}, we display a typical time evolution of the dominant wavelength extracted from the correlation function. In inset, a typical fusion event is displayed to illustrate the coarsening scenario. Because of the competition between unstable modes and the coarsening process taking place, the identification of a generic scenario for the transverse instability is problematic. This is the reason why, in addition to the experiments started from a flat bed we just described, we performed series of experiments starting from a modulated initial condition. The modulation at a given wavelength is simply produced by imprinting on the sediment surface regularly spaced thin scarifications. We find that the forced modes always fade away in region $II$, but on the other hand, in region $III$, the front modulations amplifies exponentially for a wide band of modes. The linear regime is clearly evidenced over one decade in amplitude. Non-linear effects start being visible when the amplitude becomes centimetric. The inset of figure~\ref{instalin} shows the dispersion relationship deduced from 
these experiments, which demonstrates the existence of an initial long wavelength linear instability.

\begin{figure}[t!]
\includegraphics{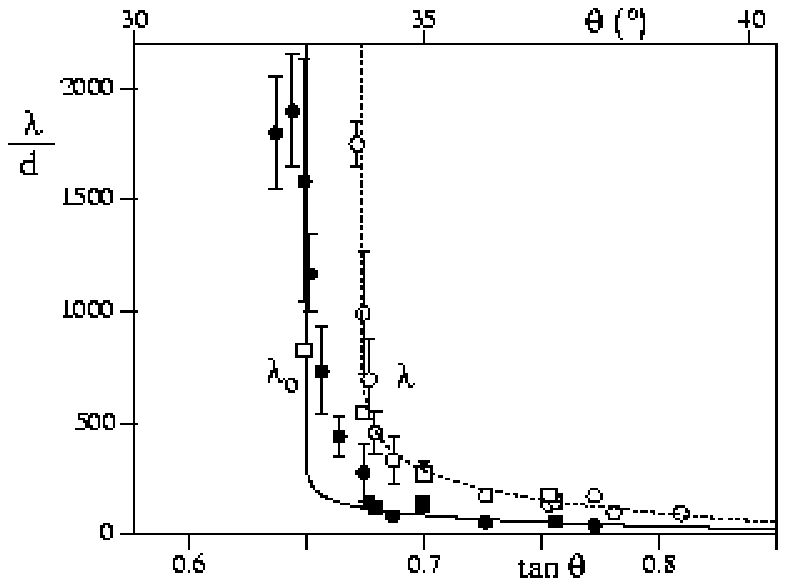}
\caption{Initial  ({\Large $\bullet$}) and final ({\Large $\circ$}) wavelengths rescaled by $d$ as a function of $\theta$ ($d=40~\mu$m, in water). In air, the wavelength obtained from the growth rate maximum ($\blacksquare$) coincides with the initial wavelength measured in water  ({\Large $\bullet$}). The final wavelength data in air ($\square$) and water  ({\Large $\circ$}) also superpose. The error bars correspond to the dispersion of the data from a realisation to the other. As $\lambda_0$ diverges at $\theta_a$, we have superimposed the curve $10~h_{start}(\theta)$ (solid line), which is a good approximation of $\lambda_0$ to the first order. The dotted line is the best fit of the final wavelength $\lambda_\infty$ by the same logarithmic form as $h_{start}(\theta)$ or $h_{stop}(\theta)$.}
\label{lambda}
\end{figure}

For experiments both in the air and under water performed in the unstable regime, we extract the two characteristic wavelengths. The initial wave length $\lambda _0$ would correspond, to the best of our experimental possibilities, to the fastest growing mode of the linear regime. Then, the wave length $\lambda _{\infty }$ is taken at the onset of the fingering instability. In fig.~\ref{lambda}, we display both wavelengths rescaled by the grains sizes :  $\lambda _{0}/d$ and $\lambda_{\infty }/d$, as a function of the inclination angle $\theta$. The selected wavelengths are typically larger than a grain size by at least two orders of magnitude. Note that the largest wavelengths measured are of the order of the track width ($1800~d$ in water and $750~d$ in air). Furthermore, in the limit of finite size effects and measurements uncertainties, we find that a value $\theta \cong \theta_{a}$ corresponds to a diverging boundary for the initial wavelength $\lambda _0/d$. It suggests that $\lambda _0$ could directly scale on $h_{start}$ (solid line in fig.~\ref{lambda}). Another striking feature is the collapse, \emph{on the same curve,} of all data both in the air, underwater and for grains of different size. In the range of parameters where the fingering regime is reached before the end of the track, the ratio of the final to the initial wavelength is approximately constant and equal to $\lambda _\infty / \lambda _{0} \simeq 3.5$. The presence of a fingering instability is a quite fascinating feature of this avalanching process. Here, the fingering front stems from the onset of localized propagating waves following the transverse instability regime. These fingers are localized matter droplets with levees on the side and propagating in a quasi solitary mode and when they are fully developed, their selected width is found to be quite sensitive to the slope (approximatively $\lambda _0$, for both underwater and dry situations).

In this letter, we have investigated the dynamics of underwater and dry granular avalanches taking place on a erodible substrate. We have identified the domain of existence for solitary waves going down the slope without changing form. For angles larger than the avalanche angle, we proved the existence of a linear transverse instability which further develops via a coarsening fusion process and finally ends up as a fingering pattern. The existence of solitary waves whose properties depend on the angle provides a new important test to models. For instance, it is easy to show that they cannot be recovered in Saint-Venant models like \cite{P99} which do not include a static erodible layer below the avalanche. The relationship between $h$ and $v$ provides an indirect evidence that a static layer does exist close to the front and below the tail of the avalanche. The mechanism responsible for the instability remains yet to be identified. Nevertheless, the inhibition of this instability on a solid bottom as well as the scaling of the initial wavelength $\lambda_{0}$ on $h_{start}$ suggests that erosion/deposition processes in the avalanche depth play a determinant role. Solving the problem experimentally is difficult as it would require an experimental access to the jammed and rolling heights in order to determine transverse modulations. In the final stage of the instability, fingers appear as droplet like solutions of the erosion/deposition process and thus look essentially different from the segregation fingers reported on a rough substrate \cite{PDS97}. Note that their shape is reminiscent of many natural patterns obtained in debris or mud flows \cite {GT04} which also display surprinsigly well selected widths at values about hundreds of a typical rock size. 

We thank P.~Claudin, O.~Pouliquen, A.~Daerr  and I.~Aranson for many discussions and suggestions on the experiment.

\end{document}